\documentclass[journal]{IEEEtran}

\usepackage[utf8]{inputenc}
\usepackage[T1]{fontenc}
\usepackage{amsmath,amssymb,bm}
\usepackage{graphicx}
\usepackage{booktabs}
\usepackage{cite}
\usepackage{tikz}
\usetikzlibrary{arrows.meta,positioning,patterns,calc}
\usepackage{pgfplots}
\pgfplotsset{compat=1.18}
\usepackage{standalone}
\usepackage[hidelinks]{hyperref}

\newcommand{\eps}{\varepsilon}
\newcommand{\kmn}{\mathbf{k}_{mn}}
\newcommand{\rv}{\mathbf{r}}
\newcommand{\kv}{\mathbf{k}}
\newcommand{\rhov}{\bm{\rho}}
\newcommand{\Jn}{J}          
\newcommand{\jn}{j}          
\newcommand{\Gtilde}{\widetilde{G}}
\newcommand{\Cs}{\mathbf{C}}  
\newcommand{\Ms}{\mathbf{M}}  

\begin{document}

\title{Analytical Modeling of the Admittance \\ Tensor of Periodic Elliptical Meta-Atoms}

\author{Thomas~Froment\`eze%
  \thanks{T.~Froment\`eze is with the University of Limoges, CNRS, XLIM, UMR 7252, F-87000 Limoges, France, and also with the Institut universitaire de France (IUF) (e-mail: thomas.fromenteze@unilim.fr).}%
  \thanks{This work was supported by the Institut universitaire de France (IUF) and the CNRS MITI ``Interactions Complexes'' program.}%
}

\maketitle

\begin{abstract}
A metasurface is an electrically thin interface that controls the phase, amplitude and polarization of a wave, upon reflection or transmission. Printed on a grounded dielectric substrate and spatially modulated across the aperture, it converts a guided surface wave into a radiated leaky wave, a mechanism used in holographic metasurface antennas. Controlling this conversion requires synthesizing a prescribed distribution of admittance tensors, each derived from the homogenization of a subwavelength unit cell. Many numerical and semi-analytical methods have been proposed to relate the geometry of a meta-atom to the tensor it produces. This paper presents, to the author's knowledge, the first analytical model, restricted to elliptical meta-atoms on a uniform lattice. The Galerkin method is applied directly to the geometry of the patches, without spatial meshing. An affine transformation maps any rotated ellipse onto the unit disk, so that a single formulation covers every configuration. A wideband response takes from minutes to tens of minutes with a periodic full-wave solver, while the proposed model evaluates it in a few tens of milliseconds on the same workstation, roughly three orders of magnitude faster. The predicted admittance agrees with a method-of-moments reference up to the fundamental resonance and captures the guided-mode anomaly of the substrate. Beyond the physical insight it brings, the model is a differentiable function of the geometry, which makes the quasi-real-time design of optimized metasurfaces possible.
\end{abstract}


\section{Introduction}

\IEEEPARstart{P}{eriodic} arrays of metallic patches on grounded dielectric substrates are widely used to design metasurfaces and holographic screens~\cite{fong2010,patel2013,faenzi2019metasurface,smith2017}. In the subwavelength regime, the array is characterized by an effective sheet admittance tensor $\mathbf{Y}_s$ relating the average tangential electric field to the induced surface current, $\mathbf{J}_s = \mathbf{Y}_s \cdot \mathbf{E}_t$~\cite{kuester2003}. In the low-frequency limit, the admittance of isolated metal elements is purely capacitive, $\mathbf{Y}_s(\omega)=j\omega\Cs$, where the symmetric positive-definite tensor $\Cs\in\mathbb{R}^{2\times2}$ accounts for patch dimensions, orientation, lattice periods, and substrate loading.

Anisotropic cells are of particular interest in this context. An elliptical patch with semi-axes $A_e\neq B_e$, rotated by an angle $\psi$, allows independent control of the principal admittance eigenvalues and their orientation. In contrast to slotted patches or double $\pi$ structures, the elliptical contour avoids sharp reentrant corners and offers smooth geometric tuning~\cite{mencagli2015,faenzi2019metasurface}. Synthesizing such metasurfaces requires determining the local geometry that achieves a prescribed admittance profile across the aperture. Solving this inverse problem can involve thousands of evaluations of the forward model, which scales the optimization time of a metasurface from seconds to hours.

\begin{figure*}[!t]
  \centering
  \includegraphics[width=\linewidth]{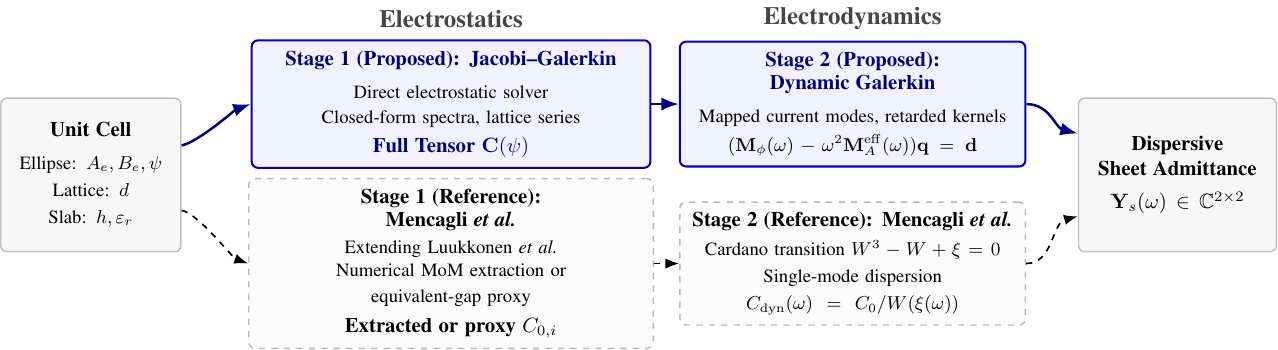}
  \caption{Two-stage analytical procedure for elliptical meta-atoms.
  Stage~1 (electrostatics): the Jacobi--Galerkin solver gives the quasi-static
  capacitance tensor $\Cs$ from the geometry, instead of a full-wave extraction or an equivalent-gap extension of planar grid models.
  Stage~2 (electrodynamics): the spectra of the charge-carrying and solenoidal surface currents define an effective magnetic interaction matrix, and the dynamic Galerkin system gives
  the dispersive sheet admittance tensor $\mathbf{Y}_s(\omega)$, to be compared with the
  single-mode Cardano model of Mencagli \emph{et al.}~\cite{mencagli2016}.}
  \label{fig:pipeline}
\end{figure*}

Determining the forward electromagnetic response involves balancing accuracy and computational cost. Analytical grid-impedance models proposed by Luukkonen \emph{et al.}~\cite{luukonen2009} yield closed-form expressions for planar strips or rectangular patches using averaged boundary conditions, but do not apply to curved boundaries. Multimodal equivalent-circuit models~\cite{rodriguezberral2015,mesa2015} extend this description to the dynamic regime by retaining a few Floquet harmonics, with a current profile on the scatterer that is fixed a priori or extracted from a single full-wave simulation. Pole-zero matching models~\cite{maci2005} describe the frequency dependence of the sheet admittance, following Foster's reactance theorem. At the other end of this range, full-wave numerical solvers handle arbitrary geometries, but their computational burden remains prohibitive for iterative design sweeps.

For elliptical geometries, Mencagli \emph{et al.}~\cite{mencagli2015} proposed a reduced spectral Method of Moments with three analytical basis functions obtained by stretching circular-patch modes. It describes the anisotropic surface-wave dispersion in the homogenization regime, well below the patch resonance, but it still requires the solution of a spectral MoM system over the Floquet lattice. In a subsequent work~\cite{mencagli2016}, the same authors proposed a two-stage semi-analytical model in which a Cardano transition function relates the dynamic surface reactance to the inductance of the grounded slab and to a single parameter per eigenaxis, the quasi-static patch capacitance~$C_0$, which allows a fast evaluation of the dispersion curve. This approach has two limitations. First, there is no analytical formulation of~$C_0$ for elliptical patches, which has to be extracted from full-wave simulations or from an extension of Luukkonen's planar-grid formula with an area-equivalent rectangular gap~\cite{luukonen2009}. Second, the single-mode approximation loses accuracy for elongated patches and departs from full-wave results well before the patch resonance.

In this context, this paper proposes an analytical spectral framework that bypasses full-wave simulations for periodic elliptical meta-atoms while maintaining wideband accuracy (Fig.~\ref{fig:pipeline}). Rather than a spatial mesh of the patch, the Galerkin method is formulated directly on the geometry. Following the disk-to-ellipse stretching similarly used in~\cite{mencagli2015,bodehou2019elliptical}, the family of rotated ellipses is projected onto a canonical unit disk, where a tailored Jacobi basis is defined. In the first stage, the quasi-static capacitance tensor $\Cs$ is extracted on this disk by expanding the surface charge on Jacobi polynomials that incorporate the Meixner edge singularity~\cite{meixner1972behavior}. With Fourier transforms available in closed form, $\Cs$ is evaluated without spatial meshing. The periodic coupling is decomposed into an isolated-patch term and a multipole series over neighboring cells, which eliminates Floquet truncations in electrostatics and extends methods known for circular disks~\cite{galindo1977,araki1981,chewkong1980}. In the second stage, this solver is coupled to a mixed-potential integral equation~\cite{michalski1997}, with surface currents transported from the disk via affine mapping~\cite{graglia1997}. This yields the dispersive admittance tensor up to the patch resonance and accounts for substrate guided modes. Because every step is expressed in closed form from the geometric parameters, the entire pipeline is a smooth function of $(A_e, B_e, \psi)$ away from the guided-mode poles, which makes it possible to compute gradients with respect to the geometry for gradient-based metasurface synthesis~\cite{budhu2022, pestourie2023physics}. A Python implementation is available as supplementary material to facilitate the reproduction of this work~\cite{code_papier}.

\section{Problem Formulation}

\begin{figure}[!b]
  \centering
  \includegraphics[width=1\columnwidth]{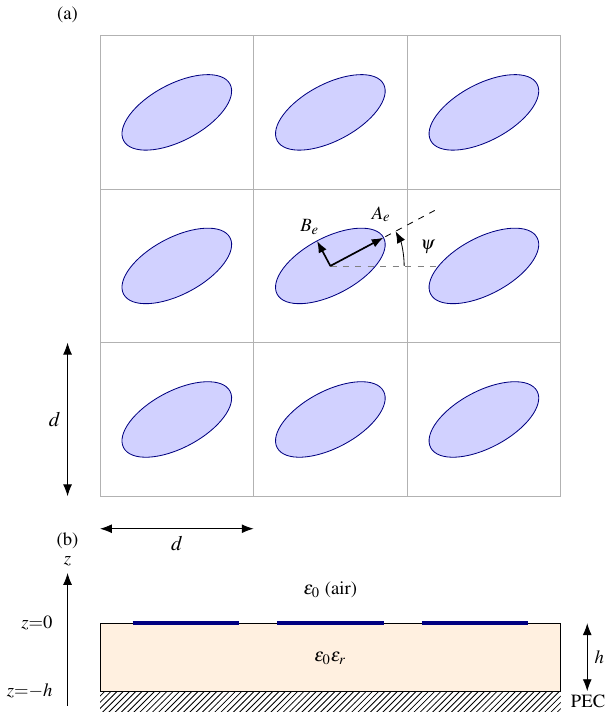}
  \caption{Problem geometry. (a)~Top view: periodic lattice (periods
  $d_x$, $d_y$) of elliptical PEC patches with semi-axes $A_e$, $B_e$
  and in-plane rotation $\psi$. (b)~Cross-section: patches at the
  air--substrate interface $z=0$, single-layer substrate
  ($\eps_r$, thickness $h$), PEC ground plane at $z=-h$.}
  \label{fig:geom}
\end{figure}

Consider a periodic array (periods $d_x$, $d_y$) of elliptical patches (semi-axes $A_e$, $B_e$, in-plane rotation $\psi$) printed on a dielectric substrate ($\eps_r$, thickness $h$) backed by a ground plane, as depicted in Fig.~\ref{fig:geom}. The metallization is modeled as a perfect electric conductor (PEC) of negligible thickness, and the dielectric substrate is assumed lossless ($\tan\delta = 0$, $\varepsilon_r \in \mathbb{R}$). For a square lattice ($d_x = d_y = d$) aligned with the patch principal axes ($\psi = 0^\circ$), the edge-to-edge gap between adjacent patches is $g = d - 2A_e$ along the major axis (and $g_y = d - 2B_e$ along the minor axis). In the quasi-static limit ($\omega\to 0$), the sheet admittance tensor reduces to
\begin{equation}
  \mathbf{Y}_s(\omega) = j\omega\,\Cs, \qquad \Cs\in\mathbb{R}^{2\times 2}.
\end{equation}

\begin{figure*}[t]
  \centering
  \includegraphics[width=\textwidth]{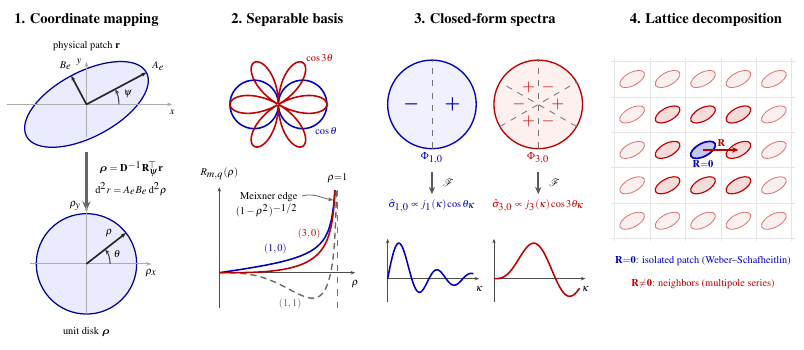}
  \caption{Four-stage analytical spectral Galerkin formulation:
  (1)~Coordinate mapping of the tilted elliptical patch to the unit disk.
  (2)~Separable basis factorized into angular harmonics $\cos m\theta$ and radial Jacobi polynomials $R_{m,q}(\rho)$ incorporating the $(1-\rho^2)^{-1/2}$ Meixner edge condition.
  (3)~Closed-form 2-D spatial modes $\Phi_{m,q}$ and their analytical Fourier spectra $\hat{\sigma}_{m,q}(\kv) \propto j_{m+2q}(\kappa)\cos m\theta_\kappa$.
  (4)~Lattice decomposition into the isolated-patch term ($\mathbf{R}=\mathbf{0}$) and a multipole series over the neighboring cells ($\mathbf{R}\neq\mathbf{0}$).}
  \label{fig:spectral_solver}
\end{figure*}

Throughout the paper, $j=\sqrt{-1}$ under the $e^{j\omega t}$ convention, $\jn_\nu$ is the spherical Bessel function of order $\nu$ and $\Jn_\nu$ the cylindrical one.
Determining $\Cs$ requires solving an electrostatic integral equation for the unknown surface charge density $\sigma(\rv)$. Periodic over the unit cell of area $S=d_x d_y$, $\sigma(\rv)$ satisfies on the patch
\begin{equation}
  \frac{1}{S}\sum_{\kmn\neq\mathbf{0}}
    \Gtilde(|\kmn|)\,\hat{\sigma}(\kmn)\,e^{j\kmn\cdot\rv}
  = V_0 + \mathbf{E}_0\cdot\rv
  \label{eq:ie}
\end{equation}
where $\kmn = (2\pi m/d_x,\, 2\pi n/d_y)$ is the Floquet harmonic, $\hat{\sigma}(\kmn)$ is the Floquet coefficient of the charge, and $\mathbf{E}_0$ is the applied uniform electric field. The $\mathbf{k}_{00}$ term drops out because each patch is electrically neutral ($\hat{\sigma}(\mathbf{0}) = 0$). Here, $\mathbf{E}_0$ denotes the macroscopic tangential field at the sheet~\cite{kuester2003,tretyakov2003}. Since all retained basis functions have an odd angular order (Section~\ref{sec:capacitance}), the constant potential $V_0$ is orthogonal to the basis and drops out of the Galerkin system.
The substrate and ground plane enter through the quasi-static spectral Green's kernel of Mosig~\cite{mosig1988} (Appendix~\ref{app:green}):
\begin{equation}
  \Gtilde(k) = \frac{1}{\eps_0\,k\,\bigl(1 + \eps_r\coth(kh)\bigr)}.
  \label{eq:Gtilde}
\end{equation}
For $kh\gg1$, $\coth(kh)\to1$ (semi-infinite substrate), whereas for $kh\ll1$, $\Gtilde\to h/(\eps_0\eps_r)$ (parallel-plate limit). The formulation can be extended to multilayer substrates with transmission-line theory, and to skewed lattices by changing the reciprocal basis that generates the harmonics $\kmn$.

\section{Spectral Galerkin Solver}
\label{sec:galerkin}

The Galerkin method expands the unknown charge as $\sigma(\rv)=\sum_n q_n\,\Phi_n(\rv)$ and tests the residual with the same functions. On a periodic lattice, this testing is carried out in the Fourier domain through Parseval's identity (Section~\ref{sec:capacitance}). Rather than a spatial mesh of the metallic patch, the Galerkin scheme is formulated directly on the geometry: an affine coordinate transformation projects the continuous elliptical domain onto a canonical unit disk, where an edge-adapted spectral basis is defined analytically (Fig.~\ref{fig:spectral_solver}).

\subsection{Coordinate Mapping to the Unit Disk}

The physical ellipse is mapped to the unit disk via the coordinate
transformation $\rv = \mathbf{R}_\psi\,\mathbf{D}\,\rhov$:
\begin{equation}
  \mathbf{D} = \operatorname{diag}(A_e,B_e),
  \qquad
  \mathbf{R}_\psi = \begin{pmatrix}\cos\psi&-\sin\psi\\\sin\psi&\cos\psi\end{pmatrix}.
\end{equation}
Points $\rhov = (\rho\cos\theta, \rho\sin\theta)$ on the unit disk ($\rho\le 1$)
map to points $\rv = (x,y)$ on an ellipse of semi-axes $A_e$ and $B_e$ rotated by an
angle $\psi$. The Jacobian of the transformation is constant and equals the area factor $A_e B_e$.

\subsection{Jacobi Basis and Edge Condition}

When an isolated conducting patch is placed in an external electric field, mutual repulsion drives the free charges toward the perimeter, where the surface charge density diverges as $(1-\rho^2)^{-1/2}$ on the mapped unit disk (Meixner edge condition~\cite{meixner1972behavior}). An expansion basis for the charge therefore has to meet three constraints: carry this edge divergence, be regular at the patch center, where a mode of angular order $m$ behaves as $\rho^m$, and resolve the smooth interior distortions caused by the neighboring cells. A basis of smooth polynomials meets the last two constraints, but not the first one: reconstructing the edge accumulation from smooth functions requires many terms and converges slowly~\cite{mittra1972analytical, lerer1993full}. The singularity is thus built into the basis itself, and only the smooth remainder has to be expanded:
\begin{equation}
  \Phi^{c/s}_{m,q}(\rho,\theta)
  = \frac{\rho^m(1-\rho^2)^{-1/2}P_q^{(-1/2,\,m)}(2\rho^2{-}1)}{N_{m,q}}
    {\textstyle\begin{Bmatrix}\cos m\theta\\[-1pt]\sin m\theta\end{Bmatrix}},
  \label{eq:basis}
\end{equation}
where $m\in\{1,3,5,\ldots,m_{\max}\}$ denotes the odd angular orders, $q\in\{0,\ldots,q_{\max}\}$ the radial indices, $N_{m,q}$ the modal normalization known in closed form~\cite{szego1939} (Appendix~\ref{app:jacobi}), and $R_{m,q}(\rho) = \rho^m(1-\rho^2)^{-1/2}P_q^{(-1/2,\,m)}(2\rho^2{-}1)/N_{m,q}$ the radial profile of the mode. The Jacobi polynomials $P_q^{(-1/2,\,m)}$ are orthogonal under the edge weight itself, and the radial index $q$ thus only describes the smooth interior of the charge.

For the fundamental dipole mode ($m=1, q=0$), the basis function reproduces the exact electrostatic charge distribution of an isolated conducting ellipse under a uniform field~\cite{landau}.
The expansion thus starts from the dominant response of an isolated patch, and the higher orders ($m\ge 3, q\ge 1$) only describe the coupling between elements, which gives fast convergence. This justifies a small basis: the six functions $(m,q)\in\{(1,0),(1,1),(3,0)\}$ of both the cosine and the sine families, $N_b=6$, are used in the rest of the paper. The adequacy of this truncation is quantified in Section~\ref{sec:benchmark_C0}, where the capacitance solver is compared with a converged reference of 50 functions ($m_{\max}=9$, $q_{\max}=4$).

\subsection{Closed-Form Fourier Spectra}

The Galerkin matrix elements require the two-dimensional Fourier transform of the charge basis functions over the patch:
\begin{equation}
  \hat{\sigma}(\kv) = \iint_{\mathrm{patch}} \sigma(\rv)\,e^{-j\kv\cdot\rv}\,d^2r.
\end{equation}
Applying the coordinate mapping $\rv = \mathbf{R}_\psi\,\mathbf{D}\,\rhov$ scales the surface element by the Jacobian $A_e B_e$.
With the kernel plane wave expanded into cylindrical harmonics (the Jacobi--Anger expansion~\cite{watson1944}, detailed in Appendix~\ref{app:jacobi}), the angular integration isolates the angular order $m$ of the mode and leaves a one-dimensional radial transform:
\begin{equation}
  \hat{\sigma}^c_{m,q}(\kv)
  = 2\pi A_eB_e\,(-j)^m\,\hat{F}_{m,q}(\kappa)\,\cos(m\theta_\kappa),
  \label{eq:sigma_spec}
\end{equation}
where $\bm{\kappa} = \mathbf{D}\mathbf{R}_\psi^T\kv = \kappa(\cos\theta_\kappa, \sin\theta_\kappa)$ is the Floquet wavevector in the disk frame, with $\sin(m\theta_\kappa)$ replacing $\cos(m\theta_\kappa)$ for the sine family $\Phi^s_{m,q}$.

The radial spectrum evaluates in closed form at every radial order $q$:
\begin{equation}
  \hat{F}_{m,q}(\kappa)
  = (-1)^q\binom{2q}{q}4^{-q}\;\frac{\jn_{m+2q}(\kappa)}{N_{m,q}},
  \label{eq:Fmq_closed}
\end{equation}
where $\jn_{m+2q}$ is the spherical Bessel function of order $m+2q$. Physically, the sharp $(1-\rho^2)^{-1/2}$ edge accumulation governs the high-wavenumber decay of the patch spectrum, which maps the Jacobi basis directly onto spherical Bessel functions in the Fourier domain~\cite{galindo1977,rahmatsamii1987} (Appendix~\ref{app:jacobi}). Both the spectra and the modal norms are obtained in closed form, and no numerical integration over the patch is needed. The remaining lattice sum of the Galerkin matrix is treated in Section~\ref{sec:lattice}.

\subsection{Capacitance Tensor Evaluation}
\label{sec:capacitance}

Projecting the electrostatic boundary condition~\eqref{eq:ie} onto the Jacobi basis functions gives the linear system
\begin{equation}
  \Ms\,\mathbf{q}=\mathbf{d}_\mathrm{phys},
\end{equation}
whose matrix elements are evaluated in reciprocal space by Parseval's identity,
\begin{equation}
  M_{ab} = \frac{1}{S}
    \sum_{\kmn\neq\mathbf{0}}
    \Gtilde(|\kmn|)\,
    \bigl[\hat{\sigma}_a(\kmn)\bigr]^*\hat{\sigma}_b(\kmn).
  \label{eq:M_mat}
\end{equation}
Since $\Gtilde(k)>0$, the matrix $\Ms$ is symmetric positive-definite. The excitation vector $\mathbf{d}_\mathrm{phys} \in \mathbb{R}^{N_b \times 2}$ collects the projections of the uniform fields $\mathbf{E}_0 = \hat{x}$ and $\mathbf{E}_0 = \hat{y}$ onto the basis:
\begin{equation}
  d_{i,b} = \iint_\mathrm{patch} r_i\,\Phi_b(\rv)\,d^2r.
  \label{eq:rhs}
\end{equation}
A uniform field along a principal axis only shifts the charge toward one side of the patch, which is the pattern of the dipole mode. In the ellipse-aligned frame $\rv' = \mathbf{R}_\psi^T\rv = (A_e\rho\cos\theta, B_e\rho\sin\theta)$, the angular integral in~\eqref{eq:rhs} selects the order $m=1$, and the orthogonality of the Jacobi polynomials cancels all radial orders $q\ge 1$. For each polarization, $\mathbf{d}_\mathrm{phys}$ has therefore a single non-zero entry, the dipole mode $(m,q)=(1,0)$ of the cosine or of the sine family, equal to $2\pi A_e^2B_e/\sqrt{6}$ and $2\pi A_eB_e^2/\sqrt{6}$ respectively (Appendix~\ref{app:jacobi}), and rotated by $\psi$ in the lattice frame. The other modes are excited only through the coupling with the neighboring cells in $\Ms$.
The lattice is centrosymmetric, and the spectrum of a mode of order $m$ changes by the factor $(-1)^m$ under $\kv\to-\kv$, so that the terms of~\eqref{eq:M_mat} cancel when $m_a+m_b$ is odd. Even and odd orders are thus not coupled, and the even orders are left out of the basis for any lattice and any rotation~$\psi$.

Under an applied field $\mathbf{E}_0$, the induced dipole moment is $\mathbf{p} = \mathbf{d}_\mathrm{phys}^T\mathbf{q}$.
The macroscopic surface capacitance tensor $\Cs\in\mathbb{R}^{2\times 2}$, relating the average surface polarization density $\mathbf{P} = \mathbf{p}/S$ to $\mathbf{E}_0$, follows directly:
\begin{equation}
  \Cs = \frac{1}{S}\,\mathbf{d}_\mathrm{phys}^T\,\Ms^{-1}\,\mathbf{d}_\mathrm{phys}.
\end{equation}

\subsection{Lattice Sum Without Floquet Truncation}
\label{sec:lattice}

Evaluating the Galerkin matrix directly in the spectral domain via~\eqref{eq:M_mat} converges slowly. The kernel~\eqref{eq:Gtilde} decays as $1/k$, and each spectrum~\eqref{eq:Fmq_closed}, proportional to a spherical Bessel function, as $1/\kappa$. The summands of~\eqref{eq:M_mat}, products of the kernel and of two spectra, therefore fall off as $|\kmn|^{-3}$ on average, and cutting off the sum at $|\kmn|\le\Lambda$ leaves a truncation error of order $1/\Lambda$. Physically, this slow spectral convergence reflects the sharp localized nature of the near-field interaction around each patch. It is therefore of interest to reformulate the interaction in real space, where the ground plane screens the fields of distant cells. Because all retained basis functions have an odd angular order, $\hat{\sigma}_a(\mathbf{0})=0$, and the harmonic $\mathbf{k}_{00}$ can be added to~\eqref{eq:M_mat} without altering the sum. Applying the Poisson summation formula then transforms the reciprocal-space series into a sum of pair interactions $W_{ab}(\mathbf{R})$ over the direct lattice translations $\mathbf{R}$:
\begin{align}
  M_{ab} &= \sum_{\mathbf{R}} W_{ab}(\mathbf{R}), \nonumber\\
  W_{ab}(\mathbf{R}) &= \int \frac{d^2k}{(2\pi)^2}\,\Gtilde(|\kv|)\,\bigl[\hat{\sigma}_a(\kv)\bigr]^*\hat{\sigma}_b(\kv)\,e^{j\kv\cdot\mathbf{R}}.
  \label{eq:poisson}
\end{align}
This decomposition separates the problem into the self-interaction of an isolated patch on the grounded slab ($\mathbf{R}=\mathbf{0}$) and the mutual coupling with neighboring cells ($\mathbf{R}\neq\mathbf{0}$).

To evaluate $\mathbf{R}=\mathbf{0}$, the quasi-static kernel is expanded into a series of ground-plane images (Appendix~\ref{app:green}, \eqref{eq:images}). The free-space self-interaction ($n=0$) evaluates in closed form through Weber--Schafheitlin integrals~\cite{araki1981}. The substrate image contributions ($n \ge 1$) are exponentially damped by the round-trip distance $2nh$ across the slab. Their remaining integrals are smooth one-dimensional spectral profiles, evaluated by exponentially convergent Gauss--Laguerre rules and in elementary closed form for the dominant dipole mode, so that the evaluation never resorts to a spatial mesh (Appendix~\ref{app:jacobi}).

For $\mathbf{R}\neq\mathbf{0}$, adjacent patches are disjoint. Under the ground-plane screening, a charge and its images form a neutral stack, so that the real-space kernel of~\eqref{eq:G_real} applied over the footprint of a neighboring cell behaves as that of a vertical dipole, smooth and decaying as $|\mathbf{R}|^{-3}$ (Appendix~\ref{app:green}). Expanding this smooth interaction around each lattice vector, or equivalently the product of the two modal spectra into a Taylor series around $\kv=\mathbf{0}$ in which each power $k_x^ik_y^j$ acts on the phase $e^{j\kv\cdot\mathbf{R}}$ as the derivative $(-j\partial_x)^i(-j\partial_y)^j$, turns the mutual coupling into a convergent multipole series:
\begin{equation}
  \sum_{\mathbf{R}\neq\mathbf{0}}W_{ab}(\mathbf{R}) = \sum_{i+j\le K}p^{ab}_{ij}\,(-j)^{i+j}\,\sum_{\mathbf{R}\neq\mathbf{0}}\partial_{x}^{\,i}\partial_{y}^{\,j}G(\mathbf{R}).
  \label{eq:neighbors}
\end{equation}
In this expression, the multipole moments $p^{ab}_{ij}$ are exact polynomials in $(A_e, B_e, \cos\psi, \sin\psi)$ obtained algebraically from the Bessel power series of the disk modes, whereas the lattice sums $\sum_{\mathbf{R}\neq\mathbf{0}}\partial^{\,i}_{x}\partial^{\,j}_{y}G(\mathbf{R})$ depend only on the lattice periods and the substrate. The patch geometry is thus completely decoupled from the lattice summation, which makes it possible to precompute the lattice sums once and evaluate new patch geometries through simple polynomial contractions. These terms generalize the interaction constants of dipole-array models~\cite{tretyakov2003} to an arbitrary multipole order~$K$, and the convergence with respect to $K$ is quantified in Section~\ref{sec:benchmark_C0}. The Taylor expansion converges when the disks circumscribing neighboring patches do not overlap ($2A_e < \min_{\mathbf{R}\neq\mathbf{0}}|\mathbf{R}|$), a condition stricter than ellipse non-overlap when $\psi\neq 0$ and satisfied by all the configurations of this paper. Details are given in Appendix~\ref{app:jacobi}.

\section{Inductive Extension}

\subsection{High-Frequency Inductive Effects}
\label{sec:inductive_effects}

The quasi-static model describes the capacitive fringing fields, but the patch also carries surface currents that store magnetic energy. This inductive contribution is negligible at low frequency and progressively compensates the capacitive susceptance, which gives a series resonance at a frequency~$f_\mathrm{res}$ where $\mathrm{Im}(Y_s) \to \infty$.

The Cardano model of Mencagli \emph{et al.}~\cite{mencagli2016} describes the grounded substrate by an equivalent unit-cell inductance $L_\mathrm{eq} = (\varepsilon_r - 1)\mu_0 h / \varepsilon_r$. Together with the quasi-static capacitance $C_{0,i}$ along the principal axis $i \in \{1,2\}$ and the displacement capacitance of the slab $\varepsilon_0\varepsilon_r h / 3$, it defines a reference frequency:
\begin{equation}
  f_{\mathrm{eq},i} = \frac{1}{2\pi\sqrt{L_\mathrm{eq}\left(C_{0,i} + \frac{1}{3}\varepsilon_0\varepsilon_r h\right)}}.
  \label{eq:feq}
\end{equation}
The dynamic admittance of the Cardano model diverges at $f_{\mathrm{eq},i}$, and the benchmarks adopt an empirical validity bound at $0.5\,f_{\mathrm{eq},i}$, where its predictions depart from the full-wave reference. The physical resonance of the patch currents differs from this reference on elongated patches, whose longer major-axis current path stores more magnetic energy, and predicting it requires the dynamic current distribution and the retarded vector potential.

\subsection{Mixed-Potential Integral Equation}

At normal incidence, the vanishing tangential electric field on the patch gives the mixed-potential integral equation:
\begin{equation}
  \mathbf{E}_0 - \nabla_s\phi(\mathbf{r}) - j\omega\,\mathbf{A}_t(\mathbf{r}) = \mathbf{0},
  \label{eq:efie_mp}
\end{equation}
where $\phi$ is the dynamic scalar potential produced by the surface charge $\sigma$ and $\mathbf{A}_t$ is the in-plane vector potential produced by the surface current $\mathbf{J}$.
In the Floquet domain, each harmonic $\kmn$ couples through the dynamic kernels $\widetilde{G}_\phi(k,\omega)$ and $\widetilde{G}_A(k,\omega)$ of the grounded slab (Formulation~C of Michalski and Mosig~\cite{michalski1997}, Appendix~\ref{app:green}), which include the retardation and the multiple reflections in the substrate and tend to the static kernels~\eqref{eq:Gtilde} when $\omega\to 0$.

The average of the charge over the cell vanishes, $\hat{\sigma}(\mathbf{0})=0$, but the average of the current does not, $\hat{\mathbf{J}}(\mathbf{0})\neq \mathbf{0}$. The harmonic $\mathbf{k}_{00}$ is excluded from the dynamic potentials in~\eqref{eq:efie_mp}, because the TEM propagation and the reflection on the ground plane are already included in the transmission-line model in which the sheet is embedded.

The surface current on the patch is decomposed into charge-carrying parts (gradients on the unit disk, mapped to the ellipse) and divergence-free (solenoidal) circulating parts. Divided by $j\omega$, the expansion reads:
\begin{equation}
  \frac{\mathbf{J}(\rv)}{j\omega} = \sum_a q_a\,\mathbf{J}_{g,a}(\rv) + \sum_b c_b\,\mathbf{J}_{s,b}(\rv),
\end{equation}
where the charge-carrying currents $\mathbf{J}_{g,a}$ satisfy $\nabla\cdot\mathbf{J}_{g,a} = -\sigma_a$, which ensures charge conservation for each electrostatic mode. The solenoidal currents $\mathbf{J}_{s,b}$ are divergence-free ($\nabla\cdot\mathbf{J}_{s,b}=0$) and have no normal flux across the patch contour ($\mathbf{J}_{s,b}\cdot\hat{\mathbf{n}} = 0$).
Testing~\eqref{eq:efie_mp} with the current basis functions turns the scalar potential into the electric interaction matrix and the vector potential term $j\omega\mathbf{A}_t$ into an $\omega^2$ inductive term, which gives the block linear system:
\begin{equation}
  \begin{bmatrix}
    \mathbf{M}_\phi(\omega) - \omega^2\,\mathbf{M}_{A,gg}(\omega) & -\omega^2\,\mathbf{M}_{A,gs}(\omega) \\[4pt]
    -\omega^2\,\mathbf{M}_{A,sg}(\omega) & -\omega^2\,\mathbf{M}_{A,ss}(\omega)
  \end{bmatrix}
  \begin{bmatrix}
    \mathbf{q} \\
    \mathbf{c}
  \end{bmatrix}
  =
  \begin{bmatrix}
    \mathbf{d} \\
    \mathbf{0}
  \end{bmatrix},
  \label{eq:YLC}
\end{equation}
where $\mathbf{M}_\phi(\omega)$ is the electric interaction matrix evaluated with the dynamic scalar kernel $\widetilde{G}_\phi(k,\omega)$, $\mathbf{d}\equiv\mathbf{d}_\mathrm{phys}$ is the incident projection matrix~\eqref{eq:rhs}, and the magnetic interaction blocks are defined by:
\begin{equation}
  M_{A,\mu\nu,ab}(\omega) = \tfrac{1}{S} \sum_{\kmn\!\neq\!\mathbf{0}}\!\widetilde{G}_A(|\kmn|,\!\omega)[\hat{\mathbf{J}}_{\mu,a}(\kmn)]^*\!\cdot\!\hat{\mathbf{J}}_{\nu,b}(\kmn)\!,
  \label{eq:MA}
\end{equation}
for $\mu,\nu \in \{g, s\}$.
Because solenoidal modes carry no charge, they couple neither to the scalar potential nor to the uniform incident field ($\mathbf{d}_s = \mathbf{0}$). The second row of~\eqref{eq:YLC} shows that the circulating currents are purely induced by the charge-carrying modes ($\mathbf{c} = -\mathbf{M}_{A,ss}^{-1}\mathbf{M}_{A,sg}\mathbf{q}$). Eliminating $\mathbf{c}$ folds their magnetic reaction into an effective $6\times 6$ system for the charge coefficients alone, with as many current modes as charge modes (Section~\ref{sec:current_modes}):
\begin{equation}
  \bigl[\mathbf{M}_\phi(\omega) - \omega^2\,\mathbf{M}_A^\mathrm{eff}(\omega)\bigr]\,\mathbf{q} = \mathbf{d},
  \label{eq:reduced_sys}
\end{equation}
with the effective magnetic interaction matrix:
\begin{equation}
  \mathbf{M}_A^\mathrm{eff}(\omega) = \mathbf{M}_{A,gg}(\omega) - \mathbf{M}_{A,gs}(\omega)\,\mathbf{M}_{A,ss}^{-1}(\omega)\,\mathbf{M}_{A,sg}(\omega).
\end{equation}
When $\omega\to 0$, the inductive term $\omega^2\mathbf{M}_A^\mathrm{eff}$ vanishes and $\mathbf{M}_\phi(0) = \Ms$, so that~\eqref{eq:reduced_sys} gives back the electrostatic system $\Ms\mathbf{q}=\mathbf{d}$. Solving for $\mathbf{q}(\omega)$ at each frequency gives the sheet admittance tensor:
\begin{equation}
  \mathbf{Y}_s(\omega) = \frac{j\omega}{S}\,\mathbf{d}^T\mathbf{q}(\omega).
  \label{eq:tensor}
\end{equation}
The patch resonance $f_\mathrm{res}$ is the pole of $\mathbf{Y}_s$, where the sheet changes from capacitive to inductive ($\mathrm{Im}(Z_s)=0$). In contrast with fixed-profile circuit models~\cite{rodriguezberral2015,mesa2015}, the modal weights $\mathbf{q}(\omega)$ and $\mathbf{c}(\omega)$ are solved at each frequency, and the current profile on the patch changes with frequency.

The matrices $\mathbf{M}_\phi(\omega)$ and $\mathbf{M}_{A,\mu\nu}(\omega)$ are lattice sums of the same form as~\eqref{eq:M_mat}. Their static limits are evaluated as in Section~\ref{sec:lattice}, with the kernel $\Gtilde(k)$ for the scalar potential and the magnetostatic image kernel of Appendix~\ref{app:green} for the vector potential. For the magnetic blocks, the harmonic $\mathbf{k}_{00}$ excluded in~\eqref{eq:MA} has to be removed from the Poisson sum, since the mean current does not vanish. For two charge-carrying modes it equals $\mu_0 h\,\mathbf{d}_a\!\cdot\mathbf{d}_b/S$, where $\mathbf{d}_a$ is the $a$-th row of $\mathbf{d}_\mathrm{phys}$ and $\widetilde{G}_A(0)=\mu_0 h$. By charge conservation, the spatial average of a charge-carrying current mode equals the dipole moment of its associated charge ($\hat{\mathbf{J}}_{g,a}(\mathbf{0}) = \mathbf{d}_a$), whereas solenoidal modes have zero mean current. Subtracting this $\mathbf{k}_{00}$ term simply removes the parallel-plate inductance of the grounded slab, which is already accounted for by the transmission-line section hosting the sheet.

To avoid slow frequency-by-frequency Floquet summations, the dynamic matrices are evaluated via a two-term Kummer-type asymptotic extraction (Appendix~\ref{app:green}):
\begin{align}
  \mathbf{M}_\phi(\omega) &= \Ms + k_0^2\,\mathbf{M}_\phi^{(1)} + \Delta\mathbf{M}_\phi(\omega), \label{eq:Mphi_decomp}\\
  \mathbf{M}_{A,\mu\nu}(\omega) &= \mathbf{M}_{A,\mu\nu}(0) + k_0^2\,\mathbf{M}_{A,\mu\nu}^{(1)} + \Delta\mathbf{M}_{A,\mu\nu}(\omega). \label{eq:MA_decomp}
\end{align}
Here, $\mathbf{M}_\phi^{(1)}$ and $\mathbf{M}_{A,\mu\nu}^{(1)}$ capture the leading frequency-independent retardation of the high-wavenumber Floquet tail and are precomputed once during unit-cell setup over a fixed grid ($m^2+n^2 \le 12^2$), whose convergence is quantified in Section~\ref{sec:benchmark_dynamic}. Because the remaining dynamic kernels decay rapidly as $k^{-5}$, the online residual blocks $\Delta\mathbf{M}_\phi(\omega)$ and $\Delta\mathbf{M}_{A,\mu\nu}(\omega)$ require summing over only the twelve lowest Floquet harmonics ($N_\mathrm{shell}=2$, or 6 centrosymmetric pairs).

\subsection{Current Modes on the Ellipse}
\label{sec:current_modes}

Following the quasi-Helmholtz (loop--star) decomposition~\cite{vecchi1999}, the surface current on the canonical unit disk is split into charge-carrying modes ($\mathbf{J}_{g,a}^\mathrm{disk} = \nabla_{\rhov}\chi_{0,a}$, irrotational on the disk) and divergence-free solenoidal modes ($\mathbf{J}_{s,b}^\mathrm{disk} = \hat{z} \times \nabla_{\rhov} \Psi_b$). Enforcing zero normal current on the rim ($\rho=1$) sets the boundary conditions: the potential $\chi_{0,a}$ satisfies the interior Neumann condition $\nabla^2\chi_{0,a} = -\sigma_a$ with $\partial_\rho\chi_{0,a}\big|_{\rho=1} = 0$, while the stream function $\Psi_b$ vanishes on the rim, $\Psi_b\big|_{\rho=1}=0$:
\begin{equation}
  \nabla^2\chi_{0,a} = -\sigma_a, \qquad
  \partial_\rho\chi_{0,a}\big|_{\rho=1} = 0, \qquad
  \Psi_b\big|_{\rho=1} = 0.
  \label{eq:neumann}
\end{equation}
Retaining the same six angular-radial orders as in Section~\ref{sec:galerkin} yields $N_g = 6$ charge-carrying and $N_s = 6$ solenoidal modes. Section~\ref{sec:benchmark_rotated} quantifies the convergence of this truncation. The spatial potential $\chi_{0,a}$ never needs to be solved explicitly, since applying Green's identity on the disk expresses $\hat{\mathbf{J}}_{g,a}$ in closed form from the charge spectrum $\hat{\sigma}_a$ and a rim boundary term (Appendix~\ref{app:currents}).

To transport these disk modes onto the physical patch, the current lines must follow the geometric stretching and rotation of the ellipse without creating artificial charge buildup on the rim. This is achieved through the affine transformation $\mathbf{F} = \mathbf{R}_\psi\,\mathbf{D}$:
\begin{equation}
  \frac{\mathbf{J}(\rv)}{j\omega}
  = \mathbf{R}_\psi\,\mathbf{D}\,\mathbf{J}_\mathrm{disk}(\rhov),
  \label{eq:J_affine}
\end{equation}
with $\mathbf{J}_\mathrm{disk} = \sum_a q_a \nabla_{\rhov}\chi_{0,a} + \sum_b c_b\,\hat{z}\times\nabla_{\rhov}\Psi_b$. Physically, scaling the current components by $A_e$ and $B_e$ stretches the current paths along the major axis and compresses them along the minor axis, which naturally accounts for the anisotropic magnetic energy storage discussed in Section~\ref{sec:inductive_effects}. Since $\nabla_{\rv} = \mathbf{R}_\psi\,\mathbf{D}^{-1}\nabla_{\rhov}$, this mapping preserves both the surface divergence ($\nabla_{\rv}\cdot(\mathbf{J}/j\omega) = \nabla_{\rhov}\cdot\mathbf{J}_\mathrm{disk} = -\sigma_\mathrm{disk}(\rhov)$) and the zero normal flux condition on the contour~\cite{graglia1997}, so that $\sigma(\rv) = \sigma_\mathrm{disk}(\rhov)$. Together with the surface element $d^2r = A_eB_e\,d^2\rho$, each spectral current mode on the ellipse follows as $\hat{\mathbf{J}}_{\mu,a}(\kv) = A_eB_e\,\mathbf{R}_\psi\,\mathbf{D}\,\hat{\mathbf{J}}_{\mu,a}^\mathrm{disk}(\bm{\kappa})$ (Appendix~\ref{app:currents}), and charge conservation holds by construction in the spectral domain ($j\kv\cdot\hat{\mathbf{J}}_{g,a}(\kv) = -\hat{\sigma}_a(\kv)$). The model remains a smooth function of the patch geometry.

For metasurface synthesis, it is also of interest to verify whether the principal axes of $\mathbf{Y}_s(\omega)$ rotate with frequency. On a representative patch tilted by $\psi = 30^\circ$ with $A_e/B_e = 2$, the rotation remains below $0.4^\circ$ up to $35$\,GHz relative to the static capacitance axes, so the electrostatic principal axes can be used as a fixed eigenbasis across the operating band.

\section{Numerical Validation}

\subsection{Consistency Checks}

Before the full-wave comparisons, the formulation is checked against known limits. At $\omega\to 0$, the dynamic kernels of Appendix~\ref{app:green} match the static kernel~\eqref{eq:Gtilde} with a relative difference below $10^{-10}$.

For an isolated patch in free space ($d\to\infty$, $\varepsilon_r=1$, $h\to\infty$, where $\Gtilde(k)\to 1/(2\varepsilon_0 k)$), the capacitance tensor is related to the polarizability volume $\bm{\alpha}_\mathrm{pol}$ by $\Cs = \varepsilon_0 \bm{\alpha}_\mathrm{pol}/S$. For a conducting ellipse with semi-axes $A_e > B_e$ and $e^2=1-(B_e/A_e)^2$, the Landau--Lifshitz polarizabilities along the principal axes are~\cite{landau}:
\begin{equation}
\begin{split}
  \alpha_a &= \tfrac{4\pi}{3}\frac{A_e^3 e^2}{K(e)-E(e)},\\[4pt]
  \alpha_b &= \tfrac{4\pi}{3}\frac{A_e^3 e^2}{E(e)/(1-e^2)-K(e)},
\end{split}
  \label{eq:LL}
\end{equation}
where $K(e)$ and $E(e)$ are the complete elliptic integrals of the first and second kind of modulus $e$ (parameter $m=e^2$). With $d_\mathrm{iso}=10\max(A_e,B_e)$, the solver reproduces~\eqref{eq:LL} within $0.2\,\%$, which is consistent with the residual dipolar coupling between neighboring cells at this spacing.

\begin{figure*}[!t]
  \centering
  \includegraphics[width=0.88\textwidth]{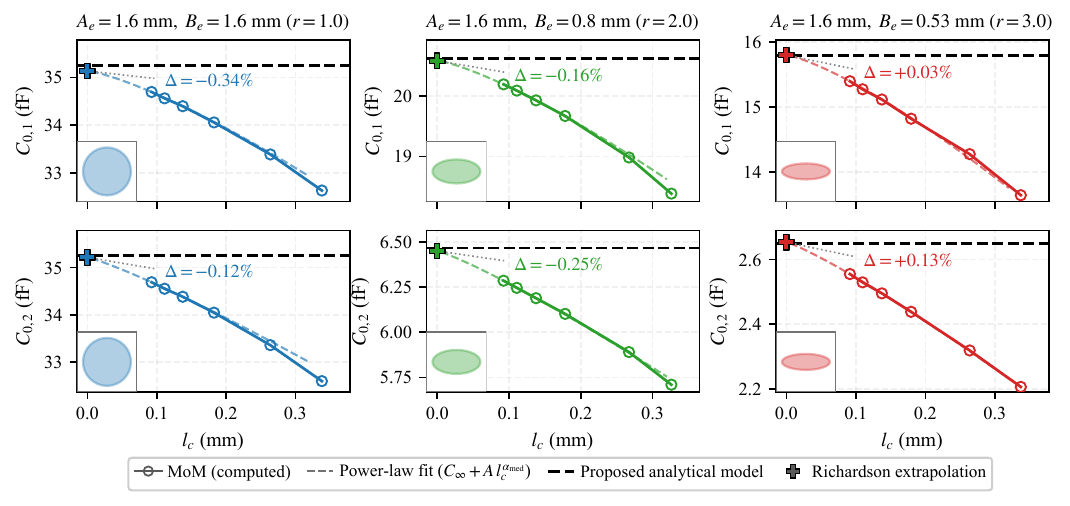}
  \caption{Convergence of the periodic MoM capacitance $C_0(l_c)$
  towards the proposed analytical Jacobi--Galerkin solver as the mesh size $l_c$ is
  refined, for three elliptical patch aspect ratios on RO4003C
  ($\varepsilon_r = 3.55$, $h = 0.813$\,mm, $d = 4$\,mm, $g/d=0.2$).
  Top row: major eigenvalue $C_{0,1}$ (power-law fitting exponent $\alpha_\mathrm{med} = 1.299$).
  Bottom row: minor eigenvalue $C_{0,2}$ ($\alpha_\mathrm{med} = 1.161$).}
  \label{fig:convergence}
\end{figure*}

\subsection{Full-Wave Periodic MoM Reference Solver}

As a reference, a full-wave periodic Method of Moments (MoM) solver was implemented for planar metallic patches on a grounded dielectric slab. The patches are meshed with triangles supporting Rao--Wilton--Glisson (RWG) basis functions~\cite{rao1982}, and the periodic Green's function of the slab is computed by a Floquet-mode summation accelerated with Kummer's transformation~\cite{singh1990,valerio2007}. The sheet admittance tensor $\mathbf{Y}_s$ is extracted from the reflection and transmission coefficients under normal incidence~\cite{kuester2003}, and the quasi-static capacitance $\Cs$ from the low-frequency slope of the susceptance ($\mathbf{Y}_s \approx j\omega\Cs$). Mesh convergence was checked with six mesh densities (from about 80 to more than 3000 unknowns), which also give the sequence used for the Richardson extrapolation.
A fine-mesh frequency sweep takes from several minutes to tens of minutes per unit cell on a standard workstation, whereas the proposed analytical model sets up a unit cell in under 25\,ms and then evaluates each frequency point in about 0.6\,ms (single core, Python implementation provided in~\cite{code_papier}), which is roughly three orders of magnitude faster than full-wave sweeps.

\subsection{Quasi-Static Capacitance Benchmark}
\label{sec:benchmark_C0}

The computational cost of Stage~1 is governed by the isolated-patch term (evaluated via analytical free-space integrals and 1-D damped image quadratures, Appendix~\ref{app:jacobi}) and by the polynomial contraction of the neighbor series~\eqref{eq:neighbors} with the precomputed lattice sums. These constants depend only on the lattice and on the substrate, and are computed once. With the $N_b=6$ basis functions and a multipole order of 30, the only truncation parameter, the capacitance tensor $\Cs$ of a new patch is obtained in approximately 25\,ms, and in about 0.6\,s on the first call when the substrate lattice sums are precomputed from scratch.

\begin{figure*}[!t]
  \centering
  \includegraphics[width=0.95\textwidth]{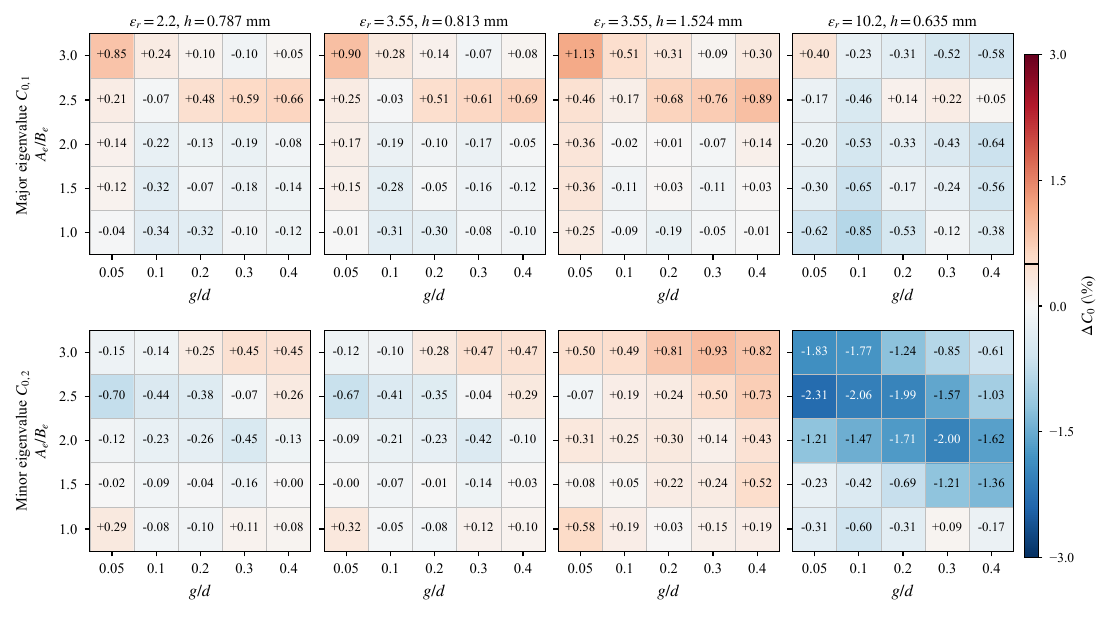}
  \caption{Relative discrepancy $\Delta C_0 = (C_{0,\mathrm{MoM}} - C_{0,\mathrm{Galerkin}})/C_{0,\mathrm{Galerkin}} \times 100\,\%$
  between the proposed analytical Jacobi--Galerkin solver and the Richardson-extrapolated MoM reference
  across 100 benchmark configurations spanning five aspect ratios, five normalized gaps,
  and four substrate configurations (RT/duroid~5880: $\varepsilon_r=2.2$, $h=0.787$\,mm; RO4003C: $\varepsilon_r=3.55$, $h=0.813$ and $1.524$\,mm; RO3010: $\varepsilon_r=10.2$, $h=0.635$\,mm). Top: major eigenvalue $C_{0,1}$. Bottom: minor eigenvalue $C_{0,2}$.
}
  \label{fig:benchmark}
\end{figure*}

The truncation to $N_b=6$ adopted in Section~\ref{sec:galerkin} is checked first. The 50 functions obtained for $m_{\max}=9$ and $q_{\max}=4$ serve as a converged reference, and the six-function subset reproduces it to within $0.5\,\%$ on $\Cs$ over the configurations of this section, with a median of $0.01\,\%$.

The MoM reference is then obtained by Richardson extrapolation (Fig.~\ref{fig:convergence}). The capacitance follows a power law $C(l_c)=C_\infty + A\,l_c^\alpha$ in the characteristic mesh size $l_c$, with a slow algebraic rate ($\alpha \approx 1.2$--$1.3$) that reflects the difficulty of piecewise-linear RWG triangles in resolving the singular edge charge. The extrapolated continuum limit $C_\infty$ matches the analytical Galerkin solver within a fraction of a percent for all three configurations.

The solver is then tested on the 100 configurations of Fig.~\ref{fig:benchmark}, a $5\times5$ grid of aspect ratios and normalized gaps spanning four substrate configurations. Among the 200 eigenvalues, $92\,\%$ agree with the reference within $\pm 1\,\%$ and $75.5\,\%$ within $\pm 0.5\,\%$. The disagreements are not distributed at random. Indeed, 15 of the 16 eigenvalues outside $\pm 1\,\%$ are minor eigenvalues $C_{0,2}$ on RO3010, the worst case reaching $-2.31\,\%$ for $A_e/B_e=2.5$ and $g/d=0.05$. Refining the model itself ($N_b$ to 50, more images, higher quadrature order and multipole order) moves these values by less than $0.01\,\%$, the multipole truncation contributing at most $0.3\,\%$ in the tightest gap $g/d=0.05$. The residual discrepancy is therefore attributed to the convergence of the MoM reference, whose uncertainty can be estimated from the circular patch, for which the two axes should coincide and differ by about $0.3\,\%$.

\subsection{Wideband Admittance Validation}
\label{sec:benchmark_dynamic}

\begin{figure*}[!t]
  \centering
  \includegraphics[width=0.95\textwidth]{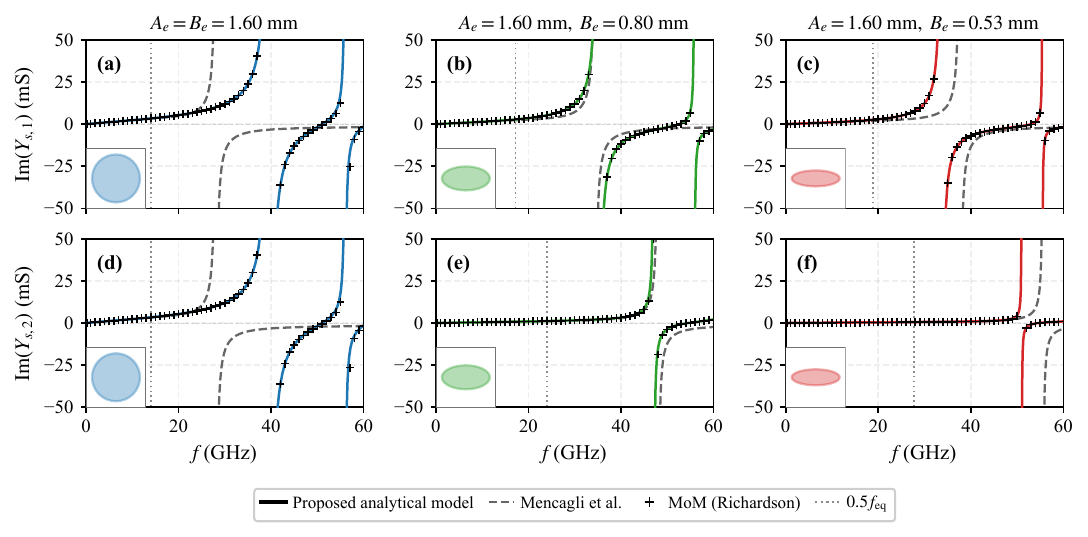}
  \caption{Dispersive sheet admittance eigenvalues $\mathrm{Im}(Y_{s,1})(f)$ (top row) and
  $\mathrm{Im}(Y_{s,2})(f)$ (bottom row) for the three representative unit-cell geometries of
  Fig.~\ref{fig:convergence} on RO4003C ($\varepsilon_r=3.55$, $h=0.813$\,mm, $d=4$\,mm, $g/d=0.2$):
  Proposed analytical model (solid colored curves~\eqref{eq:tensor}),
  Mencagli \emph{et al.} Cardano reference~\cite{mencagli2016} (dashed),
  and Richardson-extrapolated periodic MoM ($+$ markers).
  The gray dotted lines mark the reference frequency $0.5\,f_{\mathrm{eq},i}$~\eqref{eq:feq} along each principal axis. The horizontal dashed line marks
  $\mathrm{Im}(Y_s)=0$.}
  \label{fig:Y_cases}
\end{figure*}

\begin{figure*}[!t]
  \centering
  \includegraphics[width=0.95\textwidth]{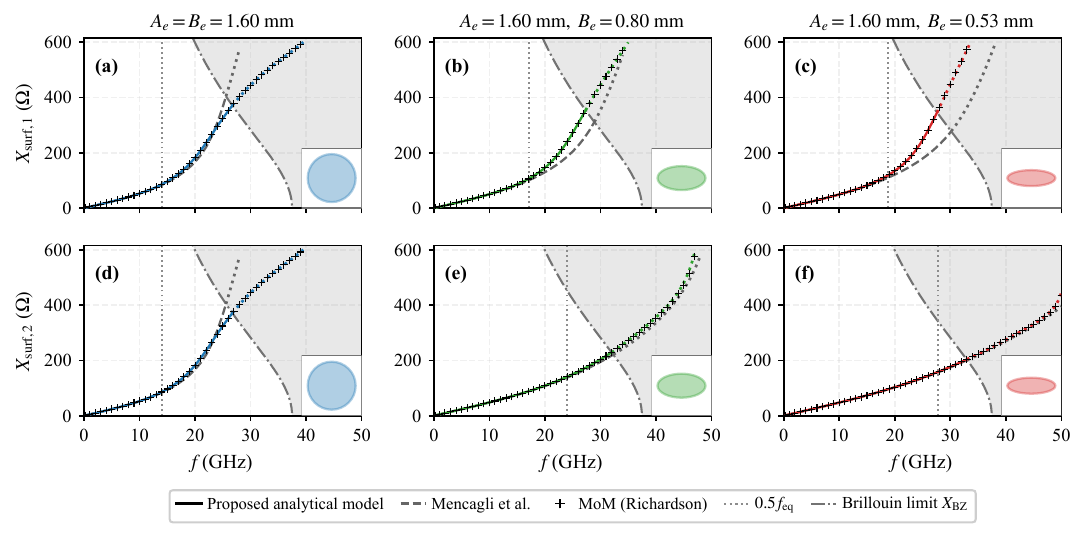}
  \caption{TM surface reactance $X_{\mathrm{surf},i}(f)$ from the transverse-resonance condition
  for the three representative geometries of Fig.~\ref{fig:Y_cases} on RO4003C
  ($\varepsilon_r=3.55$, $h=0.813$\,mm, $d=4$\,mm, $g/d=0.2$):
  (a)--(c) major axis, (d)--(f) minor axis.
  Proposed analytical model (solid colored curves), Mencagli \emph{et al.} Cardano reference~\cite{mencagli2016}
  (dashed), and Richardson-extrapolated MoM ($+$ markers).
  The dash-dotted line marks the first Brillouin zone boundary $X_\mathrm{BZ}(f)$,
  and the shaded region indicates the domain outside the first Brillouin zone ($k_\rho \ge \pi/d$,
  homogenized model not applicable), where model curves transition to dotted lines.
  The gray dotted lines mark $0.5\,f_{\mathrm{eq},i}$~\eqref{eq:feq}.}
  \label{fig:surface_reactance}
\end{figure*}

The dispersive behavior is examined on the three geometries replicated from Fig.~\ref{fig:convergence}. The eigenvalues $\mathrm{Im}(Y_{s,1})(f)$ and $\mathrm{Im}(Y_{s,2})(f)$ of the proposed model are compared in Fig.~\ref{fig:Y_cases} with the Cardano model~\cite{mencagli2016} and the MoM reference.
At low frequency, both analytical models coincide with the MoM data ($j\omega C$).
At $0.5\,f_{\mathrm{eq},i}$, the Cardano model already underestimates the MoM susceptance by $8$ to $27\,\%$, with the largest deviation along the major axis of the $A_e/B_e=3$ patch, whereas the proposed model remains within $0.6\,\%$ of the reference and follows the full-wave data in the whole pre-resonance region.

For the elongated patch ($A_e/B_e=3.0$, panels (c) and (f)), the major-axis resonance $f_{\mathrm{res},1}$ is below the reference frequency $f_{\mathrm{eq},1}=37.7$\,GHz. The proposed model places this resonant pole at 33.7\,GHz, against 33.6\,GHz for the MoM, whereas the Cardano model diverges at $f_{\mathrm{eq},1}$.
For $A_e/B_e=2.0$, the model follows the MoM data up to 60\,GHz and gives the major-axis resonance at 35.1\,GHz, where the Cardano singularity is at 34.4\,GHz.
Around 55--57\,GHz, a sharp singularity appears in both the MoM data and the model. It corresponds to the guided-mode (Wood) anomaly~\cite{hessel1965}, where the first Floquet harmonic, of wavenumber $|\mathbf{k}_{10}| = 2\pi/d$, couples to the fundamental $\mathrm{TM}_0$ surface wave of the grounded slab. For the bare slab ($h = 0.813\,\mathrm{mm}$, $\varepsilon_r = 3.55$), the $\mathrm{TM}_0$ dispersion equation $k_z \tan(k_z h) = \varepsilon_r \sqrt{k_\rho^2 - k_0^2}$, with $k_z = \sqrt{\varepsilon_r k_0^2 - k_\rho^2}$, gives $(k_\rho)_{\mathrm{TM}0} = 2\pi/d$ at $f \approx 55.1\,\mathrm{GHz}$. In the model, this interaction is carried by the first Floquet harmonics of the retardation sum. Retaining 12 harmonics ($N_\mathrm{shell}=2$, 6 centrosymmetric pairs) places the anomaly within $0.4$\,GHz of the MoM data for all three geometries, and the resonance positions of Figs.~\ref{fig:Y_cases} and~\ref{fig:Y_psi45_RO3010} directly validate this truncation. The precomputed retardation grid $m^2+n^2 \le 12^2$ was checked against $m^2+n^2 \le 24^2$, which moves the admittance eigenvalues by less than $0.003\,\%$ up to $0.95\,f_{\mathrm{res},1}$.

\subsection{Surface-Wave Reactance}

The sheet admittance is now used to compute the surface-wave reactance. For propagation along a principal axis, the wavenumber $k_\rho$ of the fundamental TM surface wave follows from the transverse-resonance condition~\cite{mencagli2016,patel2013mtt}:
\begin{equation}
  Y_\mathrm{up}^\mathrm{TM}(k_\rho, \omega) + Y_\mathrm{down}^\mathrm{TM}(k_\rho, \omega) + Y_{s, i}(\omega) = 0,
  \label{eq:transverse_resonance}
\end{equation}
where $Y_\mathrm{up}^\mathrm{TM} = j\omega\varepsilon_0 / \sqrt{k_\rho^2 - k_0^2}$ is the admittance of the air half-space and $Y_\mathrm{down}^\mathrm{TM} = -j \frac{\omega\varepsilon_0\varepsilon_r}{k_z} \cot(k_z h)$, with $k_z = \sqrt{k_0^2\varepsilon_r - k_\rho^2}$, the admittance of the grounded slab.
The TM surface reactance is $X_{\mathrm{surf}, i} = \sqrt{k_\rho^2 - k_0^2} / (\omega\varepsilon_0) = \eta_0 \sqrt{\kappa_{\mathrm{TM}, i}^2 - 1}$, with $\kappa_{\mathrm{TM}, i} = k_\rho / k_0$.

Figure~\ref{fig:surface_reactance} compares the surface reactance for the three geometries. The MoM points are obtained by solving~\eqref{eq:transverse_resonance} with the sheet admittance extracted from normal-incidence MoM simulations.
For the circular patch, both axes give the same reactance, which follows the MoM data.
When the aspect ratio increases, the surface wave becomes anisotropic: for $A_e/B_e=3.0$, the ratio $X_{\mathrm{surf},1}/X_{\mathrm{surf},2}$ is about 2.5 at 30\,GHz (and 1.8 at 25\,GHz).

Inside the first Brillouin zone ($k_\rho \le \pi/d$), the surface wave remains uncoupled from diffraction orders. With $k_\rho/k_0 = \sqrt{1 + (X_{\mathrm{surf},i}/\eta_0)^2}$, this boundary corresponds to the edge reactance
\begin{equation}
  X_\mathrm{BZ}(f) = \eta_0 \sqrt{\left(\frac{c}{2 f d}\right)^2 - 1},
\end{equation}
shown as a dash-dotted line in Fig.~\ref{fig:surface_reactance}. Beyond it, the region is shaded and curves are dotted, since the homogenized model does not describe an unattenuated guided mode there (for $d = 4$\,mm, the light line reaches the zone edge at $f = c/(2d) = 37.5$\,GHz). 

\subsection{Rotated High-Permittivity Benchmark}
\label{sec:benchmark_rotated}

To test the model in a demanding case, a patch combining several difficulties is considered: the most elongated geometry of this study ($A_e = 1.60$\,mm, $B_e = 0.53$\,mm, $A_e/B_e \approx 3.0$, $d = 4.0$\,mm, corresponding to $g/d = 0.20$ at $\psi = 0^{\circ}$) on the highest-permittivity substrate (Rogers RO3010, $\varepsilon_r = 10.2$, $h = 0.635$\,mm), rotated by $\psi = 45^\circ$. This orientation gives a strong coupling along the diagonals of the square lattice and non-zero off-diagonal entries of the tensor. The case also serves to quantify the truncation of the current basis adopted in Section~\ref{sec:current_modes}. Doubling $N_g$ and $N_s$ shifts the fundamental resonance by less than $1$\,MHz.

\begin{figure}[!t]
  \centering
  \includegraphics[width=0.88\columnwidth]{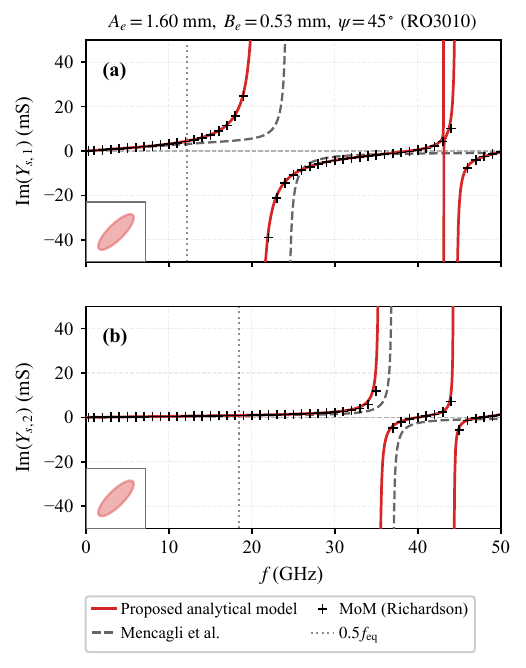}
  \caption{Dispersive sheet admittance eigenvalues for a strongly elongated patch rotated by $\psi = 45^\circ$ on high-permittivity Rogers RO3010 ($\varepsilon_r = 10.2$, $h = 0.635$\,mm, $d = 4$\,mm, $A_e = 1.60$\,mm, $B_e = 0.53$\,mm, $g/d = 0.20$ at $\psi = 0^{\circ}$): (a)~$\mathrm{Im}(Y_{s,1})(f)$, (b)~$\mathrm{Im}(Y_{s,2})(f)$. Proposed analytical model (solid red), Mencagli \emph{et al.} Cardano model~\cite{mencagli2016} using rotated static capacitance (dashed gray), and Richardson-extrapolated periodic MoM ($+$ markers). The gray dotted lines mark $0.5\,f_{\mathrm{eq},i}$~\eqref{eq:feq}. Insets in the bottom-left corner show the unit cell and the rotated patch geometry.}
  \label{fig:Y_psi45_RO3010}
\end{figure}

Figure~\ref{fig:Y_psi45_RO3010} shows the eigenvalues (a)~$\mathrm{Im}(Y_{s,1})(f)$ and (b)~$\mathrm{Im}(Y_{s,2})(f)$ up to 50\,GHz.
In panel~(a), the model places the fundamental resonance at $20.7$\,GHz, within $0.5$\,GHz of the MoM value of $20.2$\,GHz, and follows the capacitive-to-inductive transition of the reference. The higher-order substrate-coupled resonance is also reproduced, at $44.5$\,GHz against $45.1$\,GHz for the MoM.
The Cardano model~\cite{mencagli2016}, even with the exact rotated static capacitance $C_{0,1} = 40.35$\,fF, deviates from the MoM data by $30\,\%$ at $0.5\,f_{\mathrm{eq},1} = 12.16$\,GHz, diverges at $f_{\mathrm{eq},1} = 24.33$\,GHz, which is $3.6$\,GHz above the resonance of the model, and does not reproduce the higher-order resonance.
In panel~(b), the MoM gives the transverse resonance at $36.5$\,GHz, and the model at $35.3$\,GHz. The Cardano model deviates from the reference by $12\,\%$ at $0.5\,f_{\mathrm{eq},2} = 18.48$\,GHz.

\section{Conclusion}

Synthesizing a holographic metasurface means prescribing an admittance tensor for every cell of an aperture, and each prescription asks for a forward model of the local geometry. Full-wave simulation is too expensive for this inverse problem, and the semi-analytical models available for elliptical patches pay for their speed with a quasi-static capacitance extracted from full-wave data or from an equivalent rectangular gap, and with a single current mode per eigenaxis that departs from the reference well before the patch resonance. The model proposed in this paper takes the direct route: the Galerkin method is applied to the geometry of the patches themselves, without spatial meshing. The restriction to elliptical cells is what makes this possible. A rotated ellipse is an affine image of a circle, so the whole formulation lives on the canonical unit disk, with a tailored basis that absorbs the edge singularity and essentially amounts to cylindrical modes with closed-form Fourier spectra. Everything follows from these spectra: the capacitance tensor, the dispersive admittance and the surface-wave response, the lattice being reduced to an isolated-patch term and a multipole series, with no truncated Floquet summation.

Compared with a Richardson-extrapolated periodic Method of Moments, the model predicts the static capacitance within $1\,\%$ for $92\,\%$ of 200 tested eigenvalues, follows the dispersive admittance up to the fundamental resonance, including the capacitive-to-inductive transition on elongated patches, and captures the substrate guided-mode anomaly, where the single-mode Cardano reference diverges. A wideband response evaluates in a few tens of milliseconds, where the periodic full-wave solvers of FEM or spectral MoM type need minutes to tens of minutes.

These results open several practical perspectives. The admittance tensor gives the three quantities a unit cell is used for: the surface-wave dispersion, through the transverse-resonance condition, and the reflection and transmission of a normal-incidence plane wave, through a simple parallel combination of the sheet with the substrate response. Any other substrate is treated the same way, multilayer stacks included, since transfer-matrix methods keep the corresponding Green's kernels in closed form, and reflect- and transmitarray cells therefore follow without additional effort. With such evaluation times, the admittance tables of a complete aperture become routine to build, rather than requiring one full-wave simulation per cell. The model is also an explicit function of the geometry, and therefore differentiable: this makes it a natural building block for end-to-end differentiable design tools. The continuation of this work is to propose such tools to the community, so that apertures very large compared with the operating wavelength can be optimized in real time. Finally, extending this mesh-free spectral approach to the all-dielectric structures that dominate photonics is a natural perspective. It requires replacing the PEC Meixner edge basis with volumetric polarization currents, and accounting for the Mie resonances that appear when the cells approach the half-wavelength regime.

\section*{Acknowledgment}

Large language models (Claude, Anthropic) were used as auxiliary tools for cross-checking intermediate symbolic results, locating integrals and identities in the literature, code development, and language refinement. All theoretical derivations, implementations, and interpretations were developed and verified by the author, who assumes full responsibility for the contents of this paper.

\appendices

\section{Spectral Green's Kernels and Static Limit}
\label{app:green}

\subsection{Dynamic Transmission-Line Kernels}

The structure is made of air ($z>0$, permittivity $\eps_0$), a dielectric substrate ($-h<z<0$, permittivity $\eps_0\eps_r$) and a perfect electric conductor (PEC) ground plane at $z=-h$.
Following Michalski and Mosig~\cite{michalski1997}, which extends the single-layer mixed-potential formulation of Mosig~\cite{mosig1988}, the spectral kernels of the vector and scalar potentials at the interface $z=0$ are, with the $e^{j\omega t}$ convention:
\begin{align}
  \widetilde{G}_A(k,\omega) &= \frac{\mu_0}{\kappa_0+\kappa_s\coth(\kappa_s h)},
  \label{eq:GA_dyn_app}\\
  \widetilde{G}_\phi(k,\omega) &= \frac{1}{\varepsilon_0 k^2}\left[
    \frac{\kappa_0\kappa_s}{\kappa_s+\varepsilon_r\kappa_0\coth(\kappa_s h)} \right. \nonumber\\
    &\quad \left. +\,\frac{k_0^2}{\kappa_0+\kappa_s\coth(\kappa_s h)}\right],
  \label{eq:Gphi_dyn_app}
\end{align}
where $k = |\kv|$, $k_0 = \omega/c$, and $\kappa_0 = \sqrt{k^2-k_0^2}$ and $\kappa_s = \sqrt{k^2-\varepsilon_r k_0^2}$ are the vertical propagation constants in air and in the substrate. For the non-zero harmonics, $\kappa_0$ is real in the frequency range considered ($f < c/d = 75$\,GHz for $d=4$\,mm), and the substrate terms are even functions of $\kappa_s$, so that no choice of branch is needed. The denominator of~\eqref{eq:Gphi_dyn_app} has a real pole when a Floquet harmonic matches the $\mathrm{TM}_0$ guided mode of the slab (Section~\ref{sec:benchmark_dynamic}). With a lossy substrate this pole moves into the complex plane, and in the lossless case the frequency sampling avoids the exact pole.

\subsection{Quasi-Static Limit and Image Series}

When $k_0 \to 0$, $\kappa_0 \to k$ and $\kappa_s \to k$.
The vector potential kernel~\eqref{eq:GA_dyn_app} becomes
\begin{equation}
  \widetilde{G}_A(k)\big|_{k_0\to 0} = \frac{\mu_0}{k\bigl(1+\coth(kh)\bigr)} = \frac{\mu_0}{2k}\bigl(1-e^{-2kh}\bigr),
\end{equation}
which is the magnetostatic image kernel, the term $-e^{-2kh}$ being the negative current image at $z=-2h$.
In the scalar kernel~\eqref{eq:Gphi_dyn_app}, the retarded term in $k_0^2$ vanishes and the first term gives the electrostatic kernel~\eqref{eq:Gtilde} of Mosig~\cite{mosig1988}.
With $\coth(kh)=(1+e^{-2kh})/(1-e^{-2kh})$ and $\Gamma=(\eps_r-1)/(\eps_r+1)$, this kernel is expanded in a geometric series of images,
\begin{equation}
  \Gtilde(k) = \frac{1}{\eps_0(1+\eps_r)\,k}\Bigl[1+\sum_{n\ge 1}c_n\,e^{-2nkh}\Bigr],
  \label{eq:images}
\end{equation}
with $c_n = -(1+\Gamma)(-\Gamma)^{n-1}$, whose terms are the kernel of a homogeneous medium of permittivity $(1+\eps_r)/2$ and the images of the charge in the ground plane, reflected at the dielectric interface with the coefficient $\Gamma$.
Since $\sum_n c_n=-1$, the leading $1/r$ behavior cancels in the real-space kernel
\begin{equation}
  G(r) = \frac{1}{2\pi\eps_0(1+\eps_r)}\Bigl[\frac{1}{r}+\sum_{n=1}^{\infty}\frac{c_n}{\sqrt{r^2+4n^2h^2}}\Bigr],
  \label{eq:G_real}
\end{equation}
which decays as $r^{-3}$ because of the screening by the ground plane. The same transformation applied to the magnetostatic kernel gives $G_A(r)=(\mu_0/4\pi)\bigl[1/r-(r^2+4h^2)^{-1/2}\bigr]$, which does not depend on $\eps_r$.

\subsection{Asymptotic Extraction and Dynamic Convergence}

Following standard Kummer-type acceleration~\cite{singh1990,valerio2007}, the slow high-wavenumber tail of the dynamic Floquet sum is extracted analytically. For $k \gg k_0$, the dynamic kernels separate into the quasi-static kernels of the previous subsection and a frequency-independent retardation correction:
\begin{equation}
  \widetilde{G}_{\phi,A}(k,\omega) \approx \widetilde{G}_{\phi,A}^\mathrm{qs}(k) + k_0^2\,\widetilde{g}_{\phi,A}^{(1)}(k), \label{eq:asymp_kernels}
\end{equation}
where $\widetilde{g}_{\phi,A}^{(1)}(k) = \partial\widetilde{G}_{\phi,A}/\partial(k_0^2)\big|_{k_0=0}$ follows by direct differentiation of~\eqref{eq:GA_dyn_app}--\eqref{eq:Gphi_dyn_app}. This correction is pre-integrated once into the setup matrices $\mathbf{M}_\phi^{(1)}$ and $\mathbf{M}_{A,\mu\nu}^{(1)}$, over a fixed Floquet grid ($m^2+n^2 \le 12^2$). Once it is subtracted, the residual part of the sum decays fast enough that the frequency-dependent summation is restricted to the twelve lowest harmonics ($N_\mathrm{shell}=2$, or 6 centrosymmetric pairs). These harmonics include the substrate guided-mode poles ($\mathrm{TM}_0$ Wood anomalies), which are therefore still captured, in under $1$\,ms per frequency point.

\section{Jacobi Basis, Analytical Spectra, and Lattice Sum}
\label{app:jacobi}

\subsection{Modal Normalization}

Under the radial substitution $u = 2\rho^2 - 1$, the basis functions~\eqref{eq:basis} are orthogonal on the unit disk with respect to the regularizing edge weight $(1-\rho^2)^{1/2}$. Standard Jacobi polynomial orthogonality~\cite{szego1939} directly yields the closed-form modal norm:
\begin{equation}
  N_{m,q}^2 = \frac{1}{2(2q+m+1/2)}\,\frac{\Gamma(q+1/2)\,\Gamma(q+m+1)}{q!\,\Gamma(q+m+1/2)},
\end{equation}
with $N_{1,0} = \sqrt{2/3}$ ($3N_{1,0} = \sqrt{6}$) for the fundamental dipole mode.

\subsection{Analytical Fourier Spectra}

In the disk frame $\bm{\kappa} = \mathbf{D}\mathbf{R}_\psi^T\kv = \kappa(\cos\theta_\kappa, \sin\theta_\kappa)$, expanding the Fourier kernel $e^{-j\bm{\kappa}\cdot\rhov}$ into cylindrical harmonics via the Jacobi--Anger identity~\cite{watson1944} and integrating over $\theta$ selects the angular order $m$ and gives the factorized form~\eqref{eq:sigma_spec}. The remaining radial Hankel transform is a Sonine--Schafheitlin integral tabulated in~\cite{erdelyi1954} and widely used in Jacobi--Bessel aperture theory~\cite{galindo1977,rahmatsamii1987}:
\begin{multline}
  \int_0^1 x^{m+1}\,(1-x^2)^\alpha\,P_q^{(\alpha,m)}(2x^2-1)\,\Jn_m(\kappa x)\,dx\\
  = (-1)^q\,2^\alpha\,\frac{\Gamma(q+\alpha+1)}{q!}\,\kappa^{-\alpha-1}\,\Jn_{m+2q+\alpha+1}(\kappa),
  \label{eq:general_hankel}
\end{multline}
valid for $\alpha > -1$, with the argument convention $P_q^{(\alpha,m)}(2x^2{-}1)=(-1)^q\,P_q^{(m,\alpha)}(1{-}2x^2)$ relative to the tabulated form. Setting $\alpha=-1/2$ (Meixner charge weight) and using $\jn_n(\kappa) = \sqrt{\pi/(2\kappa)}\,\Jn_{n+1/2}(\kappa)$ directly gives the closed-form charge spectrum~\eqref{eq:Fmq_closed}. Setting $\alpha=+1/2$ similarly gives the solenoidal stream-function spectrum in Appendix~\ref{app:currents}.

\subsection{Isolated-Patch Term and Neighbor Series}

Because the real-space kernel~\eqref{eq:G_real} decays as $r^{-3}$ and the modal spectra are band-limited by the patch size, the Poisson lattice sum~\eqref{eq:poisson} is absolutely convergent.

\textit{Isolated-patch term ($\mathbf{R}=\mathbf{0}$).}
In the disk frame, the change of variables to $\bm{\kappa}$ gives $d^2k = d^2\kappa/(A_eB_e)$ and $|\kv|=\kappa\,g(\theta_\kappa)$, with $g(\theta_\kappa)=\sqrt{\cos^2\theta_\kappa/A_e^2+\sin^2\theta_\kappa/B_e^2}$. Substituting the image series~\eqref{eq:images} separates the self-interaction into radial and angular integrals:
\begin{multline}
  W_{ab}(\mathbf{0}) = \frac{1}{(2\pi)^2A_eB_e\,\eps_0(1+\eps_r)}\sum_{n\ge 0}c_n\int_0^{2\pi}\!\frac{d\theta_\kappa}{g(\theta_\kappa)}\\
  \times\int_0^\infty\!d\kappa\,\bigl[\hat{\sigma}_a^*\hat{\sigma}_b\bigr](\kappa,\theta_\kappa)\,e^{-2nh\,g(\theta_\kappa)\kappa},
  \qquad c_0=1.
\end{multline}
For both charge and current modes, the radial integrand reduces to a finite sum of Bessel products $\kappa^{-\gamma}\Jn_\mu(\kappa)\Jn_\nu(\kappa)$. For the free-space term ($n=0$), the radial integral over $[0,\infty)$ evaluates in closed form via the Weber--Schafheitlin integral in terms of Gamma functions~\cite[Sec.~13.41]{watson1944}, as in open circular microstrip analysis~\cite{araki1981}. For the substrate images ($n \ge 1$), the exponential damping $e^{-2nh\,g(\theta_\kappa)\kappa}$ allows a fast evaluation of the radial integral via a 32-node Gauss--Laguerre quadrature (or in elementary closed form for the dipole mode). The remaining smooth periodic integral over $\theta_\kappa\in[0,2\pi)$ is then evaluated by the trapezoidal rule.

\textit{Neighbor series ($\mathbf{R}\neq\mathbf{0}$).}
For disjoint patches, the mutual interaction is governed by the low-wavenumber behavior of the modal spectra around $\bm{\kappa}=\mathbf{0}$. Because $\jn_{m+2q}(\kappa)/\kappa^m$ and $\Jn_{m\pm 1}(\kappa)/\kappa^{m\pm 1}$ are even entire functions of $\kappa = \sqrt{\kappa_x^2+\kappa_y^2}$ and $\kappa^m e^{jm\theta_\kappa} = (\kappa_x + j\kappa_y)^m$ is a homogeneous polynomial, expanding the Bessel functions in power series~\cite{watson1944} expresses every charge and current spectrum on the unit disk as an exact bivariate Taylor series in $(\kappa_x, \kappa_y)$ with closed-form rational coefficients. Multiplying the expansions of modes $a$ and $b$ up to total degree $K$ and mapping to the lattice frame via $\bm{\kappa} = \mathbf{D}\mathbf{R}_\psi^T\kv$ gives the multipole coefficients $p_{ij}^{ab}$ in~\eqref{eq:neighbors} by binomial algebra. In~\eqref{eq:poisson}, each monomial $k_x^ik_y^j$ acts on $e^{j\kv\cdot\mathbf{R}}$ as $(-j\partial_x)^i(-j\partial_y)^j$, which transfers the derivatives onto the real-space image kernel~\eqref{eq:G_real}. The lattice derivatives are computed from the algebraic recurrence obtained by differentiating $(r^2+z^2)u^2 = 1$ on the Taylor coefficients of $u = (r^2+z^2)^{-1/2}$, and summed over 24 lattice periods with an analytical integral remainder.

\section{Current Modes and Closed-Form Spectra}
\label{app:currents}

\subsection{Charge-Carrying Currents}

For an electrostatic mode $\sigma_a(\rho,\theta) = R_{m,q}(\rho)\cos(m\theta)$, the potential in~\eqref{eq:neumann} separates as $\chi_{0,a}(\rho,\theta) = g_{m,q}(\rho)\cos(m\theta)$. Rather than solving~\eqref{eq:neumann} in space, the current spectrum $\hat{\mathbf{J}}_{g,a}^\mathrm{disk}(\bm{\kappa})$ is obtained in closed form by decomposing it into longitudinal (along $\hat{\bm{\kappa}}$) and transverse components. Because the normal current $\partial_\rho\chi_{0,a}$ vanishes on the rim, the longitudinal component is fixed directly by charge conservation ($j\bm{\kappa}\cdot\hat{\mathbf{J}}_{g,a}^\mathrm{disk}=-\hat{\sigma}_a^\mathrm{disk}$), while Green's identity on the disk shows that the transverse component stems solely from the rim value $g_{m,q}(1)$:
\begin{equation}
  \hat{\mathbf{J}}_{g,a}^\mathrm{disk}(\bm{\kappa}) = \frac{j\bm{\kappa}}{\kappa^2}\,\hat{\sigma}_a^{\mathrm{disk}}(\bm{\kappa}) + \mathbf{B}_T(\bm{\kappa}),
  \label{eq:Jg_decomp}
\end{equation}
where $\mathbf{B}_T = \mathbf{B} - \hat{\bm{\kappa}}(\hat{\bm{\kappa}}\cdot\mathbf{B})$ is the transverse projection of the contour integral $\mathbf{B}(\bm{\kappa}) = \oint_{\rho=1}\chi_{0,a}\,\hat{\mathbf{n}}\,e^{-j\bm{\kappa}\cdot\rhov}\,dl$.
By Jacobi polynomial orthogonality, the rim potential vanishes for all higher radial orders ($g_{m,q}(1) = 0$ for $q\ge1$, so that these currents are purely longitudinal) and equals $g_{m,0}(1) = \sqrt{\pi}\,\Gamma(m+1)/\bigl[2m\,N_{m,0}\,\Gamma(m+3/2)\bigr]$ for $q=0$.
For the cosine family, expanding the rim integral via the Jacobi--Anger identity gives:
\begin{align}
  B_x &= \pi g_{m,0}(1)\bigl[(-j)^{m+1}\Jn_{m+1}(\kappa)\cos(m{+}1)\theta_\kappa \nonumber\\
  &\quad + (-j)^{m-1}\Jn_{m-1}(\kappa)\cos(m{-}1)\theta_\kappa\bigr], \label{eq:Bx}
\end{align}
with $B_y$ obtained by replacing $\cos(m{\pm}1)\theta_\kappa$ with $\pm\sin(m{\pm}1)\theta_\kappa$ (and similarly for the sine family).

Although the longitudinal term in~\eqref{eq:Jg_decomp} contains an apparent $1/\kappa^2$ factor, the total spectrum $\hat{\mathbf{J}}_{g,a}^\mathrm{disk}(\bm{\kappa})$ is regular at $\bm{\kappa} = \mathbf{0}$ because the spatial current is confined to the unit disk. Specifically, for $q \ge 1$ the charge spectrum vanishes as $\kappa^{m+2q}$, whereas for $q = 0$ the apparent $\kappa^{m-1}$ divergence of the longitudinal part is canceled by the leading Taylor term of the boundary contribution $\mathbf{B}_T(\bm{\kappa})$ in~\eqref{eq:Bx}. At $\bm{\kappa}=\mathbf{0}$, only the dipole mode $(m,q)=(1,0)$ remains non-zero and gives back the static dipole moment $\mathbf{d}_a$ after applying the affine mapping~\eqref{eq:J_affine}.

\subsection{Solenoidal Currents}

The solenoidal modes $\mathbf{J}_{s,b}^\mathrm{disk} = \hat{z} \times \nabla_{\rhov}\Psi_b$ describe closed current loops circulating within the patch. Requiring zero normal flux across the boundary means that the stream function $\Psi_b$ must be constant (and thus set to zero) along the rim $\rho=1$, while its radial derivative must reproduce the $(1-\rho^2)^{-1/2}$ Meixner singularity of the tangential edge current. Both physical conditions are satisfied by choosing:
\begin{equation}
  \Psi^{c/s}_{m,q}(\rhov) = \rho^m\,(1-\rho^2)^{1/2}\,P_q^{(1/2,\,m)}(2\rho^2-1)\,{\textstyle\begin{Bmatrix}\cos m\theta\\[-1pt]\sin m\theta\end{Bmatrix}},
\end{equation}
for odd $m \in \{1, 3, \dots, m_{\max}\}$ and $q \in \{0, \dots, q_{\max}\}$. Because $\Psi_b$ vanishes on the rim, integration by parts introduces no boundary contour term and yields $\hat{\mathbf{J}}_{s,b}^\mathrm{disk}=j\,(-\kappa_y,\kappa_x)\,\hat{\Psi}_b$, where $\hat{\Psi}_b$ follows directly from~\eqref{eq:general_hankel} with $\alpha=1/2$:
\begin{equation}
  \hat{\Psi}^c_{m,q}(\bm{\kappa}) = 2\pi\,(-j)^m\cos(m\theta_\kappa)\,c_q\,\frac{\jn_{m+2q+1}(\kappa)}{\kappa},
\end{equation}
with $c_q=(-1)^q\binom{2q}{q}(2q+1)\,4^{-q}$ ($c_0=1,\,c_1=-3/2,\,c_2=15/8,\dots$), and the sine family obtained by replacing $\cos m\theta_\kappa$ with $\sin m\theta_\kappa$. As for the charge-carrying modes, the physical spectrum $\hat{\mathbf{J}}_{s,b}(\kv)$ on the rotated ellipse follows from the affine mapping~\eqref{eq:J_affine} by scaling the components by $(A_e, B_e)$ and multiplying by the Jacobian $A_e B_e$.

\bibliographystyle{IEEEtran}
\bibliography{refs_awpl_analytic}

\end{document}